\documentclass[aip,jcp,reprint,noshowkeys]{revtex4-1}
\usepackage{graphicx,dcolumn,bm,xcolor,microtype,multirow,amscd,amsmath,amssymb,amsfonts,physics,mhchem,longtable,wrapfig}

\usepackage{natbib}
\usepackage[extra]{tipa}
\AtBeginDocument{\nocite{achemso-control}}
\usepackage{mathpazo,libertine}

\usepackage{hyperref}
\hypersetup{
    colorlinks=true,
    linkcolor=blue,
    filecolor=blue,
    urlcolor=blue,
    citecolor=blue
}

\definecolor{darkgreen}{HTML}{009900}
\usepackage[normalem]{ulem}

\usepackage{hyperref}
\hypersetup{
    colorlinks=true,
    linkcolor=blue,
    filecolor=blue,
    urlcolor=blue,
    citecolor=blue
}
\newcommand{\mc}{\multicolumn}
\newcommand{\fnm}{\footnotemark}
\newcommand{\fnt}{\footnotetext}
\newcommand{\tabc}[1]{\multicolumn{1}{c}{#1}}
\newcommand{\SI}{\textcolor{blue}{supporting information}}
\newcommand{\QP}{\textsc{quantum package}}

\newcommand{\ai}[1]{\hat{a}_{#1}}
\newcommand{\aic}[1]{\hat{a}^{\dagger}_{#1}}

\newcommand{\kcal}{kcal/mol}

\newcommand{\UEG}{\text{UEG}}

\newcommand{\LDA}{\text{LDA}}
\newcommand{\PBE}{\text{PBE}}
\newcommand{\PBEUEG}{\text{PBE-UEG}}
\newcommand{\PBEot}{\text{PBEot}}

\newcommand{\CBS}{\text{CBS}}

\newcommand{\lr}{\text{lr}}
\newcommand{\sr}{\text{sr}}

\newcommand{\Ne}{N}

\newcommand{\n}[2]{n_{#1}^{#2}}
\newcommand{\Ec}{E_\text{c}}
\newcommand{\E}[2]{E_{#1}^{#2}}
\newcommand{\DE}[2]{\Delta E_{#1}^{#2}}
\newcommand{\bE}[2]{\Bar{E}_{#1}^{#2}}
\newcommand{\DbE}[2]{\Delta \Bar{E}_{#1}^{#2}}

\newcommand{\e}[2]{\varepsilon_{#1}^{#2}}
\newcommand{\be}[2]{\Bar{\varepsilon}_{#1}^{#2}}

\newcommand{\wf}[2]{\Psi_{#1}^{#2}}
\newcommand{\W}[2]{W_{#1}^{#2}}
\newcommand{\w}[2]{w_{#1}^{#2}}

\newcommand{\rsmu}[2]{\mu_{#1}^{#2}}
\newcommand{\V}[2]{V_{#1}^{#2}}
\newcommand{\SO}[2]{\phi_{#1}(\br{#2})}

\newcommand{\tX}{\text{X}}

\newcommand{\Bas}{\mathcal{B}}

\newcommand{\hT}{\Hat{T}}
\newcommand{\hWee}[1]{\Hat{W}_\text{ee}^{#1}}

\newcommand{\f}[2]{f_{#1}^{#2}}
\newcommand{\Gam}[2]{\Gamma_{#1}^{#2}}

\newcommand{\br}[1]{\mathbf{r}_{#1}}
\newcommand{\dbr}[1]{d\br{#1}}

\newcommand{\ra}{\rightarrow}

\newcommand{\tn}[2]{\tilde{n}_{#1}^{#2}}
\newcommand{\ttn}[2]{\mathring{n}_{#1}^{#2}}

\newcommand{\si}{\sigma}

\newcommand{\LCPQ}{Laboratoire de Chimie et Physique Quantiques (UMR 5626), Universit\'e de Toulouse, CNRS, UPS, France}
\newcommand{\LCT}{Laboratoire de Chimie Th\'eorique (UMR 7616), Sorbonne Universit\'e, CNRS, Paris, France}

\begin{document}	

\title{Chemically Accurate Excitation Energies With Small Basis Sets}

\author{Emmanuel Giner}
\email[Corresponding author: ]{emmanuel.giner@lct.jussieu.fr}
\affiliation{\LCT}
\author{Anthony Scemama}
\affiliation{\LCPQ}
\author{Julien Toulouse}
\affiliation{\LCT}
\author{Pierre-Fran\c{c}ois Loos}
\email[Corresponding author: ]{loos@irsamc.ups-tlse.fr}
\affiliation{\LCPQ}

\begin{abstract}
\begin{wrapfigure}[13]{o}[-1.25cm]{0.5\linewidth}
	\centering
  \includegraphics[width=\linewidth]{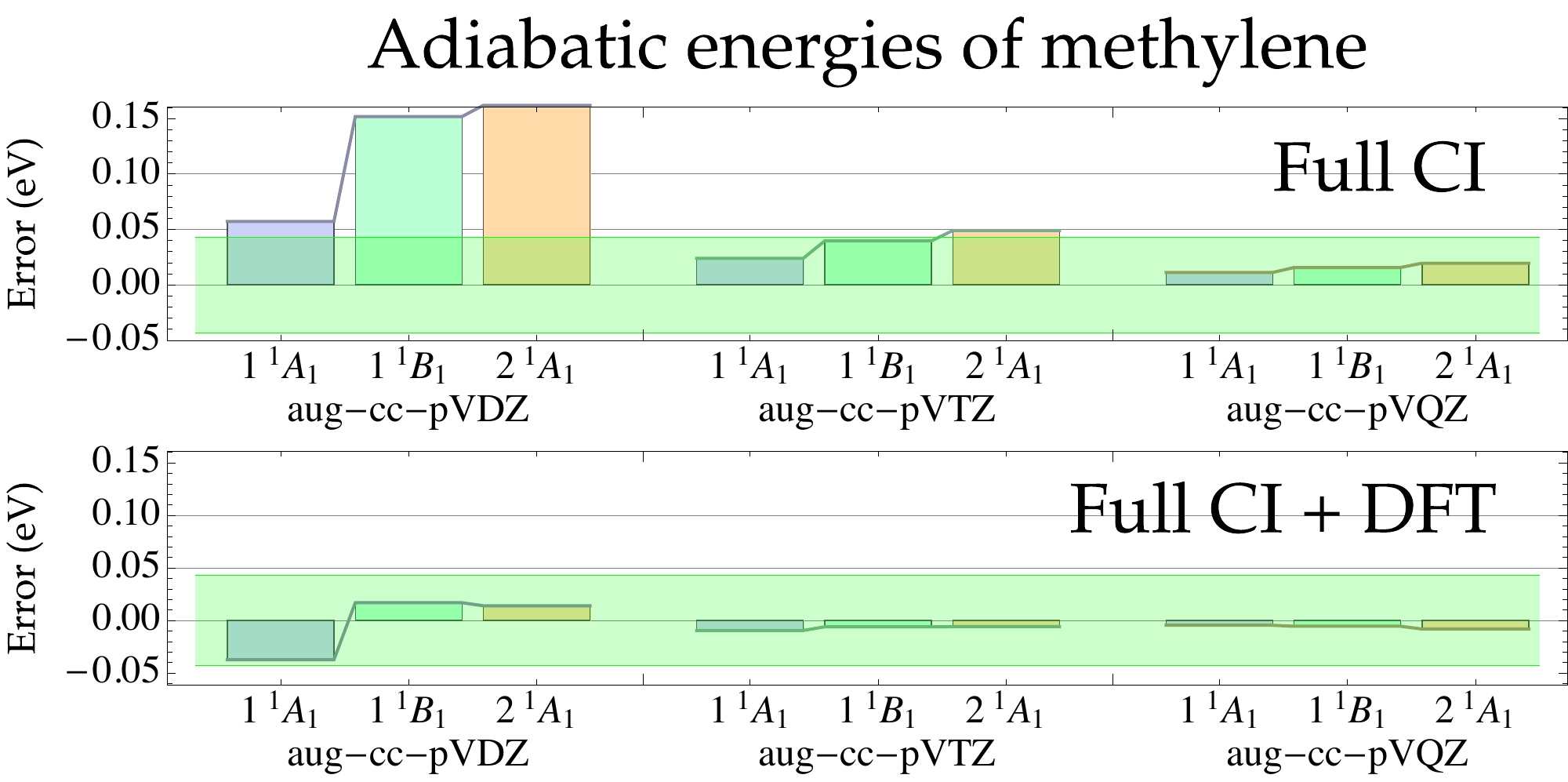}
\end{wrapfigure}
By combining extrapolated selected configuration interaction (sCI) energies obtained with the CIPSI (Configuration Interaction using a Perturbative Selection made Iteratively) algorithm with the recently proposed short-range density-functional correction for basis-set incompleteness [\href{https://doi.org/10.1063/1.5052714}{Giner \textit{et al.}, \textit{J.~Chem.~Phys.}~ \textbf{2018}, \textit{149}, 194301}], we show that one can get chemically accurate vertical and adiabatic excitation energies with, typically, augmented double-$\zeta$ basis sets.
We illustrate the present approach on various types of excited states (valence, Rydberg, and double excitations) in several small organic molecules (methylene, water, ammonia, carbon dimer and ethylene).
The present study clearly evidences that special care has to be taken with very diffuse excited states where the present correction does not catch the radial incompleteness of the one-electron basis set.
\end{abstract}

\maketitle

\section{Introduction}
\label{sec:intro}
One of the most fundamental problems of conventional wave function electronic structure methods is their slow energy convergence with respect to the size of the one-electron basis set.
The overall basis-set incompleteness error can be, qualitatively at least, split in two contributions stemming from the radial and angular incompleteness.
Although for ground-state properties angular incompleteness is by far the main source of error, it is definitely not unusual to have a significant radial incompleteness in the case of excited states (especially for Rydberg states), which can be alleviated by using additional sets of diffuse basis functions (i.e.~augmented basis sets).

Explicitly-correlated F12 methods \cite{Kut-TCA-85, Kut-TCA-85, KutKlo-JCP-91, NogKut-JCP-94} have been specifically designed to efficiently catch angular incompleteness. \cite{Ten-TCA-12, TenNog-WIREs-12, HatKloKohTew-CR-12, KonBisVal-CR-12, GruHirOhnTen-JCP-17, MaWer-WIREs-18}
Although they have been extremely successful to speed up convergence of ground-state energies and properties, such as correlation and atomization energies, \cite{TewKloNeiHat-PCCP-07} their performance for excited states \cite{FliHatKlo-JCP-06, NeiHatKlo-JCP-06, HanKoh-JCP-09, Koh-JCP-09, ShiWer-JCP-10, ShiKniWer-JCP-11, ShiWer-JCP-11, ShiWer-MP-13} has been much more conflicting. \cite{FliHatKlo-JCP-06, NeiHatKlo-JCP-06}

Instead of F12 methods, here we propose to follow a different route and investigate the performance of the recently proposed density-based basis set
incompleteness correction. \cite{GinPraFerAssSavTou-JCP-18}
Contrary to our recent study on atomization and correlation energies, \cite{LooPraSceTouGin-JPCL-19} the present contribution focuses on vertical and adiabatic excitation energies in molecular systems which is a much tougher test for the reasons mentioned above.
This density-based correction relies on short-range correlation density functionals (with multideterminant reference) from range-separated density-functional theory \cite{Sav-INC-96, LeiStoWerSav-CPL-97, TouColSav-PRA-04, TouSavFla-IJQC-04, AngGerSavTou-PRA-05, GolWerSto-PCCP-05, PazMorGorBac-PRB-06, FroTouJen-JCP-07, TouGerJanSavAng-PRL-09, JanHenScu-JCP-09, FroCimJen-PRA-10, TouZhuSavJanAng-JCP-11, MusReiAngTou-JCP-15, HedKneKieJenRei-JCP-15, HedTouJen-JCP-18, FerGinTou-JCP-19} (RS-DFT) to capture the missing part of the short-range correlation effects, a consequence of the incompleteness of the one-electron basis set.
Because RS-DFT combines rigorously density-functional theory (DFT) \cite{ParYan-BOOK-89} and wave function theory (WFT) \cite{SzaOst-BOOK-96} via a decomposition of the electron-electron interaction into a non-divergent long-range part and a (complementary) short-range part (treated with WFT and DFT, respectively), the WFT method is relieved from describing the short-range part of the correlation hole around the electron-electron coalescence points (the so-called electron-electron cusp). \cite{Kat-CPAM-57}
Consequently, the energy convergence with respect to the size of the basis set is significantly improved, \cite{FraMusLupTou-JCP-15} and chemical accuracy can be obtained even with small basis sets.
For example, in Ref.~\onlinecite{LooPraSceTouGin-JPCL-19}, we have shown that one can recover quintuple-$\zeta$ quality atomization and correlation energies with triple-$\zeta$ basis sets for a much lower computational cost than F12 methods.

This work is organized as follows. 
In Sec.~\ref{sec:theory}, the main working equations of the density-based correction are reported and discussed. 
Computational details are given in Sec.~\ref{sec:compdetails}.
In Sec.~\ref{sec:res}, we discuss our results for each system and draw our conclusions in Sec.~\ref{sec:ccl}.
Unless otherwise stated, atomic units are used.

\section{Theory}
\label{sec:theory}

The present basis-set correction assumes that we have, in a given (finite) basis set $\Bas$, the ground-state and the $k$th excited-state energies, $\E{0}{\Bas}$ and $\E{k}{\Bas}$, their one-electron densities, $\n{k}{\Bas}(\br{})$ and $\n{0}{\Bas}(\br{})$, as well as their opposite-spin on-top pair densities, $\n{2,0}{\Bas}(\br{})$ and $\n{2,k}{\Bas}(\br{})$,
Therefore, the complete-basis-set (CBS) energy of the ground and excited states may be approximated as \cite{GinPraFerAssSavTou-JCP-18}
\begin{subequations}
\begin{align}
	\label{eq:E0CBS}
	\E{0}{\CBS} & \approx \E{0}{\Bas} + \bE{}{\Bas}[\n{0}{\Bas}],
	\\
	\label{eq:EkCBS}
	\E{k}{\CBS} & \approx \E{k}{\Bas} + \bE{}{\Bas}[\n{k}{\Bas}],
\end{align}
\end{subequations}
where 
\begin{equation}
	\label{eq:E_funcbasis}
	 \bE{}{\Bas}[\n{}{}]
	= \min_{\wf{}{} \rightsquigarrow \n{}{}}  \mel*{\wf{}{}}{\hT + \hWee{}}{\wf{}{}} 
	- \min_{\wf{}{\Bas} \rightsquigarrow \n{}{}}  \mel*{\wf{}{\Bas}}{\hT + \hWee{}}{\wf{}{\Bas}}
\end{equation}
is the basis-dependent complementary density functional, 
\begin{align}
	\hT & = - \frac{1}{2} \sum_{i}^{\Ne} \nabla_i^2,
	&
	\hWee{} & = \sum_{i<j}^{\Ne} r_{ij}^{-1},
\end{align}
are the kinetic and electron-electron repulsion operators, respectively, and $\wf{}{\Bas}$ and $\wf{}{}$ are two general $\Ne$-electron normalized wave functions belonging to the Hilbert spaces spanned by $\Bas$ and the complete basis, respectively.
The notation $\wf{}{} \rightsquigarrow \n{}{}$ in Eq.~\eqref{eq:E_funcbasis} states that $\wf{}{}$ yields the one-electron density $\n{}{}$. 

Hence, the CBS excitation energy associated with the $k$th excited state reads
\begin{equation}
\begin{split}
	\DE{k}{\CBS} 
	& = \E{k}{\CBS} - \E{0}{\CBS} 
	\\
	& \approx \DE{k}{\Bas} + \DbE{}{\Bas}[\n{0}{\Bas},\n{k}{\Bas}],
\end{split}
\end{equation}
where
\begin{equation}
	\label{eq:DEB}
	\DE{k}{\Bas} = \E{k}{\Bas} - \E{0}{\Bas}
\end{equation}
is the excitation energy in $\Bas$ and
\begin{equation}
	\label{eq:DbE}
	\DbE{}{\Bas}[\n{0}{\Bas},\n{k}{\Bas}] = \bE{}{\Bas}[\n{k}{\Bas}] - \bE{}{\Bas}[\n{0}{\Bas}]
\end{equation}
its basis-set correction.
An important property of the present correction is
\begin{equation}
  \label{eq:limitfunc}
        \lim_{\Bas \to \CBS} \DbE{}{\Bas}[\n{0}{\Bas},\n{k}{\Bas}] = 0.
\end{equation}
In other words, the correction vanishes in the CBS limit, hence guaranteeing an unaltered limit. \cite{LooPraSceTouGin-JPCL-19}
Note that in Eqs.~\eqref{eq:E0CBS} and \eqref{eq:EkCBS} we have assumed that the same density functional $\bE{}{\Bas}$ can be used for correcting all excited-state energies, which seems a reasonable approximation since the electron-electron cusp effects are largely universal. \cite{Kut-TCA-85, MoKut-JPC-93, KutMor-ZPD-96, Tew-JCP-08, LooGil-MP-10, LooGil-JCP-2015}

\subsection{Range-separation function}
\label{sec:rs}

As initially proposed in Ref.~\onlinecite{GinPraFerAssSavTou-JCP-18} and further developed in Ref.~\onlinecite{LooPraSceTouGin-JPCL-19}, we have shown that one can efficiently approximate $\bE{}{\Bas}[\n{}{}]$ by short-range correlation functionals with multi-determinantal (ECMD) reference borrowed from RS-DFT. \cite{TouGorSav-TCA-05}
The ECMD functional, $\bE{\text{c,md}}{\sr}[\n{}{},\rsmu{}{}]$, is a function of the range-separation parameter $\mu$ and admits, for any $\n{}{}$, the following two limits
\begin{subequations}
\begin{align}
	\label{eq:large_mu_ecmd}
	\lim_{\mu \to \infty}  \bE{\text{c,md}}{\sr}[\n{}{},\rsmu{}{}] & = 0,
	\\
	\label{eq:small_mu_ecmd}
	\lim_{\mu \to 0}  \bE{\text{c,md}}{\sr}[\n{}{},\rsmu{}{}] & = \Ec[\n{}{}],
\end{align}
\end{subequations}
which correspond to the WFT limit ($\mu \to \infty$) and the Kohn-Sham DFT (KS-DFT) limit ($\mu = 0$).
In Eq.~\eqref{eq:small_mu_ecmd}, $\Ec[\n{}{}]$ is the usual universal correlation density functional defined in KS-DFT. \cite{HohKoh-PR-64, KohSha-PR-65}

The key ingredient that allows us to exploit ECMD functionals for correcting the basis-set incompleteness error is the range-separated function
\begin{equation}
	\label{eq:def_mu}
	\rsmu{}{\Bas}(\br{}) =  \frac{\sqrt{\pi}}{2} \W{}{\Bas}(\br{},\br{}),
\end{equation}
which automatically adapts to the spatial non-homogeneity of the basis-set incompleteness error.
It is defined such that the long-range interaction of RS-DFT, $\w{}{\lr,\mu}(r_{12}) = \erf( \mu r_{12})/r_{12}$, coincides, at coalescence, with an effective two-electron interaction $\W{}{\Bas}(\br{1},\br{2})$ ``mimicking'' the Coulomb operator in an incomplete basis $\Bas$, i.e.~$\w{}{\lr,\rsmu{}{\Bas}(\br{})}(0) = \W{}{\Bas}(\br{},\br{})$ at any $\br{}$. \cite{GinPraFerAssSavTou-JCP-18}
The explicit expression of $\W{}{\Bas}(\br{1},\br{2})$ is given by 
\begin{equation}
	\label{eq:def_weebasis}
	\W{}{\Bas}(\br{1},\br{2})   = 
 	\begin{cases}
       \f{}{\Bas}(\br{1},\br{2})/\n{2}{\Bas}(\br{1},\br{2}), 	& \text{if $\n{2}{\Bas}(\br{1},\br{2}) \ne 0$,}
       \\
       \infty, 												& \text{otherwise,}
    \end{cases}
\end{equation}
where
\begin{equation}
	\label{eq:n2basis}
	\n{2}{\Bas}(\br{1},\br{2})
	= \sum_{pqrs \in \Bas}  \SO{p}{1} \SO{q}{2} \Gam{pq}{rs} \SO{r}{1} \SO{s}{2},
\end{equation}
and $\Gam{pq}{rs} = 2 \mel*{\wf{}{\Bas}}{ \aic{r_\downarrow}\aic{s_\uparrow}\ai{q_\uparrow}\ai{p_\downarrow}}{\wf{}{\Bas}}$ are the opposite-spin pair density associated with $\wf{}{\Bas}$ and its corresponding tensor, respectively, $\SO{p}{}$ is a (real-valued) molecular orbital (MO),
\begin{equation}
	\label{eq:fbasis}
	\f{}{\Bas}(\br{1},\br{2}) 
	= \sum_{pqrstu \in \Bas}  \SO{p}{1} \SO{q}{2} \V{pq}{rs} \Gam{rs}{tu} \SO{t}{1} \SO{u}{2},
\end{equation}
and $\V{pq}{rs}= \braket{pq}{rs}$ are two-electron Coulomb integrals. 
An important feature of $\W{}{\Bas}(\br{1},\br{2})$ is that it tends to the regular Coulomb operator $r_{12}^{-1}$ as $\Bas \to \CBS${, which implies that 
\begin{equation}
	\lim_{\Bas \rightarrow \CBS} \rsmu{}{\Bas}(\br{}) = \infty,
\end{equation} 
ensuring that $\bE{}{\Bas}[\n{}{}]$ vanishes when $\Bas$ is complete.
We refer the interested readers to Refs.~\onlinecite{GinPraFerAssSavTou-JCP-18,LooPraSceTouGin-JPCL-19} for additional details.

\subsection{Short-range correlation functionals}
\label{sec:func}

The local-density approximation ($\LDA$) of the ECMD complementary functional is defined as 
\begin{equation}
	\label{eq:def_lda_tot}
	\bE{\LDA}{\Bas}[\n{}{},\rsmu{}{\Bas}] = \int \n{}{}(\br{}) \be{\text{c,md}}{\sr,\LDA}\qty(\n{}{}(\br{}),\zeta(\br{}),\rsmu{}{\Bas}(\br{})) \dbr{},
\end{equation}
where $\zeta = (\n{\uparrow}{} - \n{\downarrow}{})/\n{}{}$ is the spin polarization and $\be{\text{c,md}}{\sr,\LDA}(\n{}{},\zeta,\rsmu{}{})$ is the ECMD short-range correlation energy per electron of the uniform electron gas (UEG) \cite{LooGil-WIRES-16} parameterized in Ref.~\citenum{PazMorGorBac-PRB-06}.

The functional $\be{\text{c,md}}{\sr,\LDA}$ from Eq.~\eqref{eq:def_lda_tot} presents two main defects: i) at small $\mu$, it overestimates the correlation energy, and ii) UEG-based quantities are hardly transferable when the system becomes strongly correlated. 
An attempt to solve these problems was suggested by some of the authors in the context of RS-DFT. \cite{FerGinTou-JCP-19}  
They proposed to interpolate between the usual Perdew-Burke-Ernzerhof ($\PBE$) correlation functional \cite{PerBurErn-PRL-96} $\e{\text{c}}{\PBE}(\n{}{},s,\zeta)$ (where $s=\nabla n/n^{4/3}$ is the reduced density gradient) at $\mu = 0$ and the exact large-$\mu$ behavior. \cite{TouColSav-PRA-04, GorSav-PRA-06, PazMorGorBac-PRB-06}
In the context of RS-DFT, the large-$\mu$ behavior corresponds to an extremely short-range interaction in the short-range functional.
In this regime, the ECMD energy
\begin{align}
  \label{eq:exact_large_mu}
  \bE{\text{c,md}}{\sr} = \frac{2\sqrt{\pi} (1 - \sqrt{2})}{3\mu^3} \int \dbr{} \n{2}{}(\br{}) + \order*{\mu^{-4}}
\end{align}
only depends on the \textit{exact} on-top pair density $\n{2}{}(\br{}) \equiv \n{2}{}(\br{},\br{})$ which is obtained from the \textit{exact} ground-state wave function $\Psi$ belonging to the many-electron Hilbert space in the CBS limit.

Obviously, an exact quantity such as $\n{2}{}(\br{})$ is out of reach in practical calculations and must be approximated by a function referred here as $\tn{2}{}(\br{})$. 
For a given $\tn{2}{}(\br{})$, some of the authors proposed the following functional form in order to interpolate between $\e{\text{c}}{\PBE}(\n{}{},s,\zeta)$ at $\mu = 0$ and Eq.~\eqref{eq:exact_large_mu} as $\mu \to \infty$: \cite{FerGinTou-JCP-19} 
\begin{subequations}
\begin{gather}
	\label{eq:epsilon_cmdpbe}
	\be{\text{c,md}}{\sr,\PBE}(\n{}{},\tn{2}{},s,\zeta,\rsmu{}{}) = \frac{\e{\text{c}}{\PBE}(\n{}{},s,\zeta)}{1 + \beta^\PBE(\n{}{},\tn{2}{},s,\zeta) \rsmu{}{3} },
	\\
	\label{eq:beta_cmdpbe}
	\beta^\PBE(\n{}{},\tn{2}{},s,\zeta) = \frac{3}{2\sqrt{\pi} (1 - \sqrt{2})} \frac{\e{\text{c}}{\PBE}(\n{}{},s,\zeta)}{\tn{2}{}/\n{}{}}.
\end{gather}
\end{subequations}
As illustrated in the context of RS-DFT, \cite{FerGinTou-JCP-19} such a functional form is able to treat both weakly and strongly correlated systems thanks to the explicit inclusion of $\e{\text{c}}{\PBE}$ and $\tn{2}{}$, respectively.
Therefore, in the present context, we introduce the general form of the $\PBE$-based complementary functional within a given basis set $\Bas$ 
\begin{multline}
	\bE{\PBE}{\Bas}[\n{}{},\tn{2}{},\rsmu{}{\Bas}] =
	\int \n{}{}(\br{}) 
	\\
	\times \be{\text{c,md}}{\sr,\PBE}\qty(\n{}{}(\br{}),\tn{2}{}(\br{}),s(\br{}),\zeta(\br{}),\rsmu{}{\Bas}(\br{})) \dbr{},
\end{multline}
which has an explicit dependency on both the range-separation function $\rsmu{}{\Bas}(\br{})$ (instead of the range-separation parameter in RS-DFT) and the approximation level of $\tn{2}{}$.

In Ref.~\onlinecite{LooPraSceTouGin-JPCL-19}, some of the authors introduced a version of the $\PBE$-based functional, here-referred as $\PBEUEG$
\begin{equation}
	\bE{\PBEUEG}{\Bas} 
	\equiv
	\bE{\PBE}{\Bas}[\n{}{},\n{2}{\UEG},\rsmu{}{\Bas}],
\end{equation}
in which the on-top pair density was approximated by its UEG version, i.e., $\tn{2}{}(\br{}) = \n{2}{\UEG}(\br{})$, with
\begin{equation}
	\label{eq:n2UEG}
 \n{2}{\UEG}(\br{}) \approx n(\br{})^2 [1-\zeta(\br{})^2] g_0(n(\br{})), 
\end{equation}
and where $g_0(n)$ is the UEG on-top pair distribution function [see Eq.~(46) of Ref.~\citenum{GorSav-PRA-06}]. 
Note that in Eq.~\eqref{eq:n2UEG} the dependence on the spin polarization $\zeta$ is only approximate.
As illustrated in Ref.~\onlinecite{LooPraSceTouGin-JPCL-19}, the $\PBEUEG$ functional has clearly shown, for weakly correlated systems, to improve energetics over the pure UEG-based functional $\bE{\LDA}{\Bas}$ [see Eq.~\eqref{eq:def_lda_tot}] thanks to the leverage brought by the $\PBE$ functional in the small-$\mu$ regime.

However, the underlying UEG on-top pair density might not be suited for the treatment of excited states and/or strongly correlated systems. 
Besides, in the context of the present basis-set correction, $\n{2}{\Bas}(\br{})$, the on-top pair density in $\Bas$, must be computed anyway to obtain $\rsmu{}{\Bas}(\br{})$ [see Eqs.~\eqref{eq:def_mu} and \eqref{eq:def_weebasis}].
Therefore, as in Ref.~\onlinecite{FerGinTou-JCP-19}, we define a better approximation of the exact on-top pair density as
\begin{equation}
	\label{eq:ot-extrap}
  \ttn{2}{\Bas}(\br{}) = \n{2}{\Bas}(\br{}) \qty( 1 + \frac{2}{\sqrt{\pi}\rsmu{}{\Bas}(\br{})})^{-1}
\end{equation}
which directly follows from the large-$\mu$ extrapolation of the exact on-top pair density proposed by Gori-Giorgi and Savin \cite{GorSav-PRA-06} in the context of RS-DFT. 
Using this new ingredient, we propose here the ``$\PBE$-ontop'' (\PBEot) functional
\begin{equation}
	\bE{\PBEot}{\Bas}
	\equiv
	\bE{\PBE}{\Bas}[\n{}{},\ttn{2}{\Bas},\rsmu{}{\Bas}].
\end{equation}
The sole distinction between $\PBEUEG$ and $\PBEot$ is the level of approximation of the exact on-top pair density.

\section{Computational details}
\label{sec:compdetails}
In the present study, we compute the ground- and excited-state energies, one-electron densities and on-top pair densities with a selected configuration interaction (sCI) method known as CIPSI (Configuration Interaction using a Perturbative Selection made Iteratively). \cite{HurMalRan-JCP-73, GinSceCaf-CJC-13, GinSceCaf-JCP-15} 
Both the implementation of the CIPSI algorithm and the computational protocol for excited states is reported in Ref.~\onlinecite{SceCafBenJacLoo-RC-19}.
The total energy of each state is obtained via an efficient extrapolation procedure of the sCI energies designed to reach near-FCI accuracy. \cite{HolUmrSha-JCP-17, QP2}
These energies will be labeled exFCI in the following.
Using near-FCI excitation energies (within a given basis set) has the indisputable advantage to remove the error inherent to the WFT method.
Indeed, in the present case, the only source of error on the excitation energies is due to basis-set incompleteness.
We refer the interested reader to Refs.~\onlinecite{HolUmrSha-JCP-17, SceGarCafLoo-JCTC-18, LooSceBloGarCafJac-JCTC-18, SceBenJacCafLoo-JCP-18, LooBogSceCafJac-JCTC-19, QP2} for more details.
The one-electron densities and on-top pair densities are computed from a very large CIPSI expansion containing up to several million of Slater determinants.
All the RS-DFT and exFCI calculations have been performed with {\QP}. \cite{QP2}
For the numerical quadratures, we employ the SG-2 grid. \cite{DasHer-JCC-17}
Except for methylene for which FCI/TZVP geometries have been taken from Ref.~\onlinecite{SheLeiVanSch-JCP-98}, the other molecular geometries have been extracted from Refs.~\onlinecite{LooSceBloGarCafJac-JCTC-18, LooBogSceCafJac-JCTC-19} and have been obtained at the CC3/aug-cc-pVTZ level of theory.
For the sake of completeness, all these geometries are reported in the {\SI}.
Frozen-core calculations are systematically performed and defined as such: a \ce{He} core is frozen from \ce{Li} to \ce{Ne}, while a \ce{Ne} core is frozen from \ce{Na} to \ce{Ar}.
The frozen-core density-based correction is used consistently with the frozen-core approximation in WFT methods.
We refer the reader to Ref.~\onlinecite{LooPraSceTouGin-JPCL-19} for an explicit derivation of the equations associated with the frozen-core version of the present density-based basis-set correction.
Compared to the exFCI calculations performed to compute energies and densities, the basis-set correction represents, in any case, a marginal computational cost.
In the following, we employ the AVXZ shorthand notations for Dunning's aug-cc-pVXZ basis sets.

\section{Results and Discussion}
\label{sec:res}

\subsection{Methylene}
\label{sec:CH2}

Methylene is a paradigmatic system in electronic structure theory. \cite{Sch-Science-86} 
Due to its relative small size, its ground and excited states have been thoroughly studied with high-level ab initio methods. \cite{Sch-Science-86, BauTay-JCP-86, JenBun-JCP-88, SheVanYamSch-JMS-97, SheLeiVanSch-JCP-98, AbrShe-JCP-04, AbrShe-CPL-05, ZimTouZhaMusUmr-JCP-09, GouPieWlo-MP-10, ChiHolAdaOttUmrShaZim-JPCA-18}

As a first test of the present density-based basis-set correction, we consider the four lowest-lying states of methylene ($1\,^{3}B_1$, $1\,^{1}A_1$, $1\,^{1}B_1$ and $2\,^{1}A_1$) at their respective equilibrium geometry and compute the corresponding adiabatic transition energies for basis sets ranging from AVDZ to AVQZ.
We have also computed total energies at the exFCI/AV5Z level and used these alongside the quadruple-$\zeta$ ones to extrapolate the total energies to the CBS limit with the usual extrapolation formula \cite{HelJorOls-BOOK-02}
\begin{equation}
	\E{}{\text{AVXZ}}(\tX) = \E{}{\CBS} + \alpha \, \tX^{-3}.
\end{equation}
These results are illustrated in Fig.~\ref{fig:CH2} and reported in Table \ref{tab:CH2} alongside reference values from the literature obtained with various deterministic and stochastic approaches. \cite{ChiHolAdaOttUmrShaZim-JPCA-18, SheLeiVanSch-JCP-98, JenBun-JCP-88, SheLeiVanSch-JCP-98, ZimTouZhaMusUmr-JCP-09}
Total energies for each state can be found in the {\SI}.
The exFCI/CBS values are still off by a few tenths of a {\kcal} compared to the DMC results of Zimmerman \textit{et al.} \cite{ZimTouZhaMusUmr-JCP-09} which are extremely close from the experimentally-derived adiabatic energies.
The reason of this discrepancy is probably due to the frozen-core approximation which has been applied in our case and has shown to significantly affect adiabatic energies. \cite{LooGalJac-JPCL-18, LooJac-JCTC-19}
However, the exFCI/CBS energies are in perfect agreement with the semistochastic heat-bath CI (SHCI) calculations from Ref.~\onlinecite{ChiHolAdaOttUmrShaZim-JPCA-18}, as expected.

Figure \ref{fig:CH2} clearly shows that, for the double-$\zeta$ basis, the exFCI adiabatic energies are far from being chemically accurate with errors as high as 0.15 eV.
From the triple-$\zeta$ basis onward, the exFCI excitation energies are chemically accurate though (i.e. error below 1 {\kcal} or 0.043 eV), and converge steadily to the CBS limit when one increases the size of the basis set.
Concerning the basis-set correction, already at the double-$\zeta$ level, the $\PBEot$ correction returns chemically accurate excitation energies.
The performance of the $\PBEUEG$ and $\LDA$ functionals is less impressive. 
Yet, they still yield significant reductions of the basis-set incompleteness error, hence representing a good compromise between computational cost and accuracy.
Note that the results for the $\PBEUEG$ functional are not represented in Fig.~\ref{fig:CH2} as they are very similar to the $\LDA$ ones (similar considerations apply to the other systems studied below).
It is also quite evident that, the basis-set correction has the tendency of over-correcting the excitation energies via an over-stabilization of the excited states compared to the ground state.
This trend is quite systematic as we shall see below.

	\begin{squeezetable}
	\begin{table}
	\caption{
	Adiabatic transition energies (in eV) of excited states of methylene for various methods and basis sets.
	The relative difference with respect to the exFCI/CBS result is reported in square brackets.
	See {\SI} for total energies.}
	\label{tab:CH2}
	\begin{ruledtabular}
	\begin{tabular}{lllll}
				&				&	\mc{3}{c}{Transitions}	\\
	\cline{3-5}
	Method		&	Basis set	&	\tabc{$1\,^{3}B_1 \ra 1\,^{1}A_1$}		
								&	\tabc{$1\,^{3}B_1 \ra 1\,^{1}B_1$}	
								&	\tabc{$1\,^{3}B_1 \ra 2\,^{1}A_1$}		\\
	\hline
	exFCI		&	AVDZ					
								&	$0.441$ [$+0.057$]	
								&	$1.536$ [$+0.152$]
								&	$2.659$ [$+0.162$]	\\	
				&	AVTZ					
								&	$0.408$ [$+0.024$]	
								&	$1.423$ [$+0.040$]	
								&	$2.546$ [$+0.049$]	\\	
				&	AVQZ					
								&	$0.395$ [$+0.011$]	
								&	$1.399$ [$+0.016$]	
								&	$2.516$ [$+0.020$]	\\	
				&	AV5Z					
								&	$0.390$ [$+0.006$]	
								&	$1.392$ [$+0.008$]	
								&	$2.507$ [$+0.010$]	\\	
				&	CBS					
								&	$0.384$	
								&	$1.384$	
								&	$2.497$	\\	
								\\
	exFCI+$\PBEot$	&	AVDZ					
								&	$0.347$ [$-0.037$]	
								&	$1.401$ [$+0.017$]	
								&	$2.511$ [$+0.014$]	\\	
				&	AVTZ					
								&	$0.374$ [$-0.010$]	
								&	$1.378$ [$-0.006$]	
								&	$2.491$ [$-0.006$]	\\	
				&	AVQZ					
								&	$0.379$ [$-0.005$]	
								&	$1.378$ [$-0.006$]	
								&	$2.489$ [$-0.008$]	\\	
								\\
	exFCI+$\PBEUEG$	&	AVDZ					
								&	$0.308$ [$-0.076$]	
								&	$1.388$ [$+0.004$]	
								&	$2.560$ [$+0.064$]	\\	
				&	AVTZ					
								&	$0.356$ [$-0.028$]	
								&	$1.371$ [$-0.013$]	
								&	$2.510$ [$+0.013$]	\\	
				&	AVQZ					
								&	$0.371$ [$-0.013$]
								&	$1.375$ [$-0.009$]	
								&	$2.498$ [$+0.002$]	\\	
								\\
	exFCI+$\LDA$	&	AVDZ					
								&	$0.337$ [$-0.047$]	
								&	$1.420$ [$+0.036$]	
								&	$2.586$ [$+0.089$]	\\	
				&	AVTZ					
								&	$0.359$ [$-0.025$]	
								&	$1.374$ [$-0.010$]	
								&	$2.514$ [$+0.017$]	\\	
				&	AVQZ					
								&	$0.370$ [$-0.014$]	
								&	$1.375$ [$-0.009$]	
								&	$2.499$ [$-0.002$]	\\	
								\\
	SHCI\fnm[1]	&	AVQZ					
								&	$0.393$	
								&	$1.398$	
								&	$2.516$	\\
	CR-EOMCC (2,3)D\fnm[2]&	AV5Z					
								&	$0.430$	
								&	$1.464$	
								&	$2.633$	\\
	FCI\fnm[3]	&	TZ2P					
								&	$0.483$	
								&	$1.542$	
								&	$2.674$	\\
	DMC\fnm[4]	&							
								&	$0.406$	
								&	$1.416$	
								&	$2.524$	\\
	Exp.\fnm[5]	&							
								&	$0.406$	
								&	$1.415$	
	\end{tabular}
	\end{ruledtabular}
	\fnt[1]{Semistochastic heat-bath CI (SHCI) calculations from Ref.~\onlinecite{ChiHolAdaOttUmrShaZim-JPCA-18}.}
	\fnt[2]{Completely-renormalized equation-of-motion coupled cluster (CR-EOMCC) calculations from Refs.~\onlinecite{GouPieWlo-MP-10}.}
	\fnt[3]{Reference \onlinecite{SheLeiVanSch-JCP-98}.}
	\fnt[4]{Diffusion Monte Carlo (DMC) calculations from Ref.~\onlinecite{ZimTouZhaMusUmr-JCP-09} obtained with a CAS(6,6) trial wave function.}
	\fnt[5]{Experimentally-derived values. See footnotes of Table II from Ref.~\onlinecite{GouPieWlo-MP-10} for additional details.}
	\end{table}
	\end{squeezetable}

\begin{figure}
	\includegraphics[width=\linewidth]{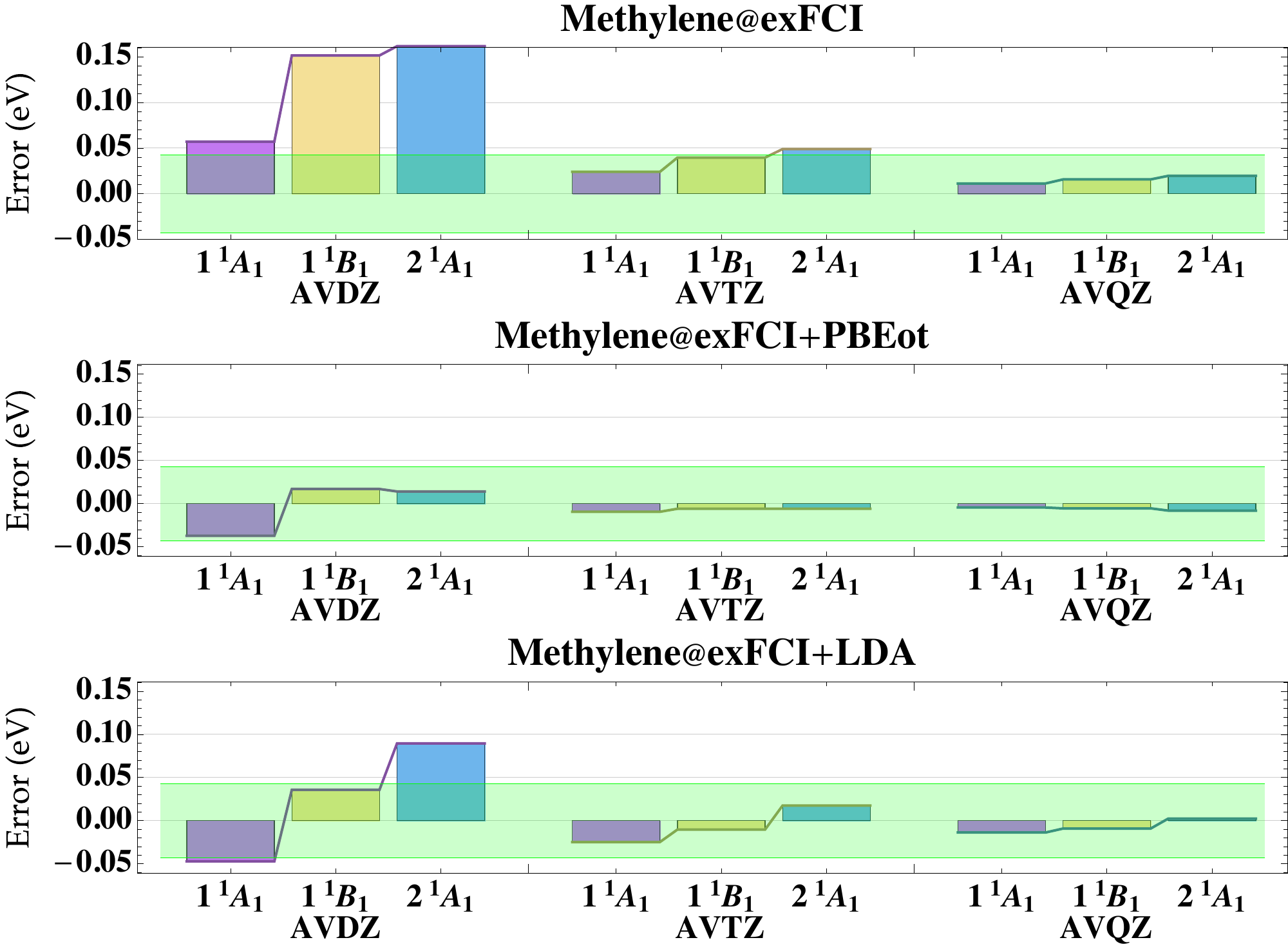}
	\caption{Error in adiabatic excitation energies (in eV) of methylene for various basis sets and methods.
	The green region corresponds to chemical accuracy (i.e., error below 1 {\kcal} or 0.043 eV).
	See Table \ref{tab:CH2} for raw data.}
	\label{fig:CH2}
\end{figure}

\subsection{Rydberg States of Water and Ammonia}
\label{sec:H2O-NH3}

For the second test, we consider the water \cite{CaiTozRei-JCP-00, RubSerMer-JCP-08, LiPal-JCP-11, LooSceBloGarCafJac-JCTC-18, SceBenJacCafLoo-JCP-18, SceCafBenJacLoo-RC-19} and ammonia \cite{SchGoe-JCTC-17, BarDelPerMat-JMS-97, LooSceBloGarCafJac-JCTC-18} molecules.
They are both well studied and possess Rydberg excited states which are highly sensitive to the radial completeness of the one-electron basis set, as evidenced in Ref.~\onlinecite{LooSceBloGarCafJac-JCTC-18}.
Table \ref{tab:Mol} reports vertical excitation energies for various singlet and triplet excited states of water and ammonia at various levels of theory (see the {\SI} for total energies).
The basis-set corrected theoretical best estimates (TBEs) have been extracted from Ref.~\onlinecite{LooSceBloGarCafJac-JCTC-18} and have been obtained on the same geometries.
These results are also depicted in Figs.~\ref{fig:H2O} and \ref{fig:NH3} for \ce{H2O} and \ce{NH3}, respectively.
One would have noticed that the basis-set effects are particularly strong for the third singlet excited state of water and the third and fourth singlet excited states of ammonia where this effect is even magnified.
In other words, substantial error remains in these cases even with the largest AVQZ basis set.
In these cases, one really needs doubly augmented basis sets to reach radial completeness.
The first observation worth reporting is that all three RS-DFT correlation functionals have very similar behaviors and they significantly reduce the error on the excitation energies for most of the states.
However, these results also clearly evidence that special care has to be taken for very diffuse excited states where the present correction cannot catch the radial incompleteness of the one-electron basis set, a feature which is far from being a cusp-related effect.

	\begin{squeezetable}
	\begin{table*}
	\caption{
	Vertical excitation energies (in eV) of excited states of water, ammonia, carbon dimer and ethylene for various methods and basis sets.
	The TBEs have been extracted from Refs.~\onlinecite{LooSceBloGarCafJac-JCTC-18, LooBogSceCafJac-JCTC-19} on the same geometries.
	See the {\SI} for total energies.}
	\label{tab:Mol}
	\begin{ruledtabular}
	\begin{tabular}{lllddddddddddddd}
				&				&					&				&	\mc{12}{c}{Deviation with respect to TBE}		
	\\
	\cline{5-16}
				&				&					&				&	\mc{3}{c}{exFCI}	
																	&	\mc{3}{c}{exFCI+$\PBEot$}	
																	&	\mc{3}{c}{exFCI+$\PBEUEG$}	
																	&	\mc{3}{c}{exFCI+$\LDA$}
	\\
	\cline{5-7}	\cline{8-10}	\cline{11-13}	\cline{14-16}
	Molecule	&	Transition		&	Nature		&	\tabc{TBE}	&	\tabc{AVDZ}	&	\tabc{AVTZ}	&	\tabc{AVQZ}	
																	&	\tabc{AVDZ}	&	\tabc{AVTZ}	&	\tabc{AVQZ}	
																	&	\tabc{AVDZ}	&	\tabc{AVTZ}	&	\tabc{AVQZ}	
																	&	\tabc{AVDZ}	&	\tabc{AVTZ}	&	\tabc{AVQZ}	
	\\
	\hline
	Water		&	$1\,^{1}A_1 \ra 1\,^{1}B_1$		&	Ryd.	&	7.70\fnm[1]		&	-0.17	&	-0.07	&	-0.02	
																	&	0.01	&	0.00	&	0.02	
																	&	-0.02	&	-0.01	&	0.00	
																	&	-0.04	&	-0.01	&	0.01
	\\
				&	$1\,^{1}A_1 \ra 1\,^{1}A_2$		&	Ryd.	&	9.47\fnm[1]		&	-0.15	&	-0.06	&	-0.01	
																	&	0.03	&	0.01	&	0.03	
																	&	0.00	&	0.00	&	0.02	
																	&	-0.03	&	0.00	&	0.00
	\\
				&	$1\,^{1}A_1 \ra 2\,^{1}A_1$		&	Ryd.	&	9.97\fnm[1]		&	-0.03	&	0.02	&	0.06	
																	&	0.13	&	0.08	&	0.09	
																	&	0.10	&	0.07	&	0.08		
																	&	0.09	&	0.07	&	0.03
	\\
				&	$1\,^{1}A_1 \ra 1\,^{3}B_1$		&	Ryd.	&	7.33\fnm[1]		&	-0.19	&	-0.08	&	-0.03	
																	&	0.02	&	0.00	&	0.02	
																	&	0.05	&	0.01	&	0.02	
																	&	0.00	&	0.00	&	0.04
	\\
				&	$1\,^{1}A_1 \ra 1\,^{3}A_2$		&	Ryd.	&	9.30\fnm[1]		&	-0.16	&	-0.06	&	-0.01	
																	&	0.04	&	0.02	&	0.04	
																	&	0.07	&	0.03	&	0.04	
																	&	0.03	&	0.03	&	0.04
	\\
				&	$1\,^{1}A_1 \ra 1\,^{3}A_1$		&	Ryd.	&	9.59\fnm[1]		&	-0.11	&	-0.05	&	-0.01	
																	&	0.07	&	0.02	&	0.03	
																	&	0.09	&	0.03	&	0.03	
																	&	0.06	&	0.03	&	0.04
	\\
	\\
	Ammonia		&	$1\,^{1}A_{1} \ra 1\,^{1}A_{2}$		&	Ryd.	&	6.66\fnm[1]	&	-0.18	&	-0.07	&	-0.04	
																				&	-0.04	&	-0.02	&	-0.01	
																				&	-0.07	&	-0.03	&	-0.02	
																				&	-0.07	&	-0.03	&	-0.02	
	\\
				&	$1\,^{1}A_{1} \ra 1\,^{1}E$		&	Ryd.		&	8.21\fnm[1]	&	-0.13	&	-0.05	&	-0.02	
																				&	0.01	&	0.00	&	0.01	
																				&	-0.03	&	-0.01	&	0.00	
																				&	-0.03	&	0.00	&	0.00	
	\\
				&	$1\,^{1}A_{1} \ra 2\,^{1}A_{1}$		&	Ryd.	&	8.65\fnm[1]	&	1.03	&	0.68	&	0.47	
																				&	1.17	&	0.73	&	0.50	
																				&	1.12	&	0.72	&	0.49	
																				&	1.11	&	0.71	&	0.49	
	\\
				&	$1\,^{1}A_{1} \ra 2\,^{1}A_{2}$		&	Ryd.	&	8.65\fnm[2]	&	1.22	&	0.77	&	0.59	
																				&	1.36	&	0.83	&	0.62	
																				&	1.33	&	0.81	&	0.61	
																				&	1.32	&	0.81	&	0.61	
	\\
				&	$1\,^{1}A_{1} \ra 1\,^{3}A_{2}$		&	Ryd.	&	9.19\fnm[1]	&	-0.18	&	-0.06	&	-0.03	
																				&	-0.03	&	0.00	&	-0.02	
																				&	-0.07	&	-0.02	&	-0.03	
																				&	-0.07	&	-0.01	&	-0.03	
	\\
	\\
	Carbon dimer	&	$1\,^{1}\Sigma_g^+ \ra 1\,^{1}\Delta_g$		&	Val.	&	2.04\fnm[3]	&	0.17	&	0.05	&	0.02
																								&	0.04	&	0.00	&	0.00
																								&	0.15	&	0.04	&	0.02
																								&	0.17	&	0.05	&	0.02
	\\
						&	$1\,^{1}\Sigma_g^+ \ra 2\,^{1}\Sigma_g^+$	&	Val.	&	2.38\fnm[3]	&	0.12	&	0.04	&	0.02
																								&	0.00	&	0.00	&	0.00
																								&	0.11	&	0.03	&	0.02
																								&	0.13	&	0.04	&	0.02
	\\
	\\
	Ethylene	&	$1\,^{1}A_{1g} \ra 1\,^{1}B_{3u}$		&	Ryd.	&	7.43\fnm[3]	&	-0.12	&	-0.04	&	
																					&	-0.05	&	-0.01	&	
																					&	-0.04	&	-0.01	&	
																					&	-0.02	&	0.00	&	
	\\
				&	$1\,^{1}A_{1g} \ra 1\,^{1}B_{1u}$		&	Val.	&	7.92\fnm[3]	&	0.01	&	0.01	&	
																					&	0.00	&	0.00	&	
																					&	0.06	&	0.03	&	
																					&	0.06	&	0.03	&	
	\\
				&	$1\,^{1}A_{1g} \ra 1\,^{1}B_{1g}$		&	Ryd.	&	8.10\fnm[3]	&	-0.1	&	-0.02	&	
																					&	-0.03	&	0.00	&	
																					&	-0.02	&	0.00	&		
																					&	0.00	&	0.01	&	
	\\
				&	$1\,^{1}A_{1g} \ra 1\,^{3}B_{1u}$		&	Val.	&	4.54\fnm[3]	&	0.01	&	0.00	&	
																					&	0.05	&	0.03	&	
																					&	0.08	&	0.04	&	
																					&	0.07	&	0.04	&	
	\\
				&	$1\,^{1}A_{1g} \ra 1\,^{3}B_{3u}$		&	Val.	&	7.28\fnm[4]	&	-0.12	&	-0.04	&	
																					&	-0.04	&	0.00	&	
																					&	0.00	&	0.00	&	
																					&	0.00	&	0.02	&	
	\\
				&	$1\,^{1}A_{1g} \ra 1\,^{3}B_{1g}$		&	Val.	&	8.00\fnm[4]	&	-0.07	&	-0.01	&	
																					&	0.00	&	0.03	&	
																					&	0.04	&	0.03	&	
																					&	0.05	&	0.04	&	
	\end{tabular}
	\end{ruledtabular}
	\fnt[1]{exFCI/AVQZ data corrected with the difference between CC3/d-AV5Z and exFCI/AVQZ values. \cite{LooSceBloGarCafJac-JCTC-18} 
	d-AV5Z is the doubly augmented V5Z basis set.}
	\fnt[2]{exFCI/AVTZ data corrected with the difference between CC3/d-AV5Z and exFCI/AVTZ values. \cite{LooSceBloGarCafJac-JCTC-18}}
	\fnt[3]{exFCI/CBS obtained from the exFCI/AVTZ and exFCI/AVQZ data of Ref.~\onlinecite{LooBogSceCafJac-JCTC-19}.}
	\fnt[4]{exFCI/AVDZ data corrected with the difference between CC3/d-AV5Z and exFCI/AVDZ values. \cite{LooSceBloGarCafJac-JCTC-18}}
	\end{table*}
	\end{squeezetable}

\begin{figure}
	\includegraphics[width=\linewidth]{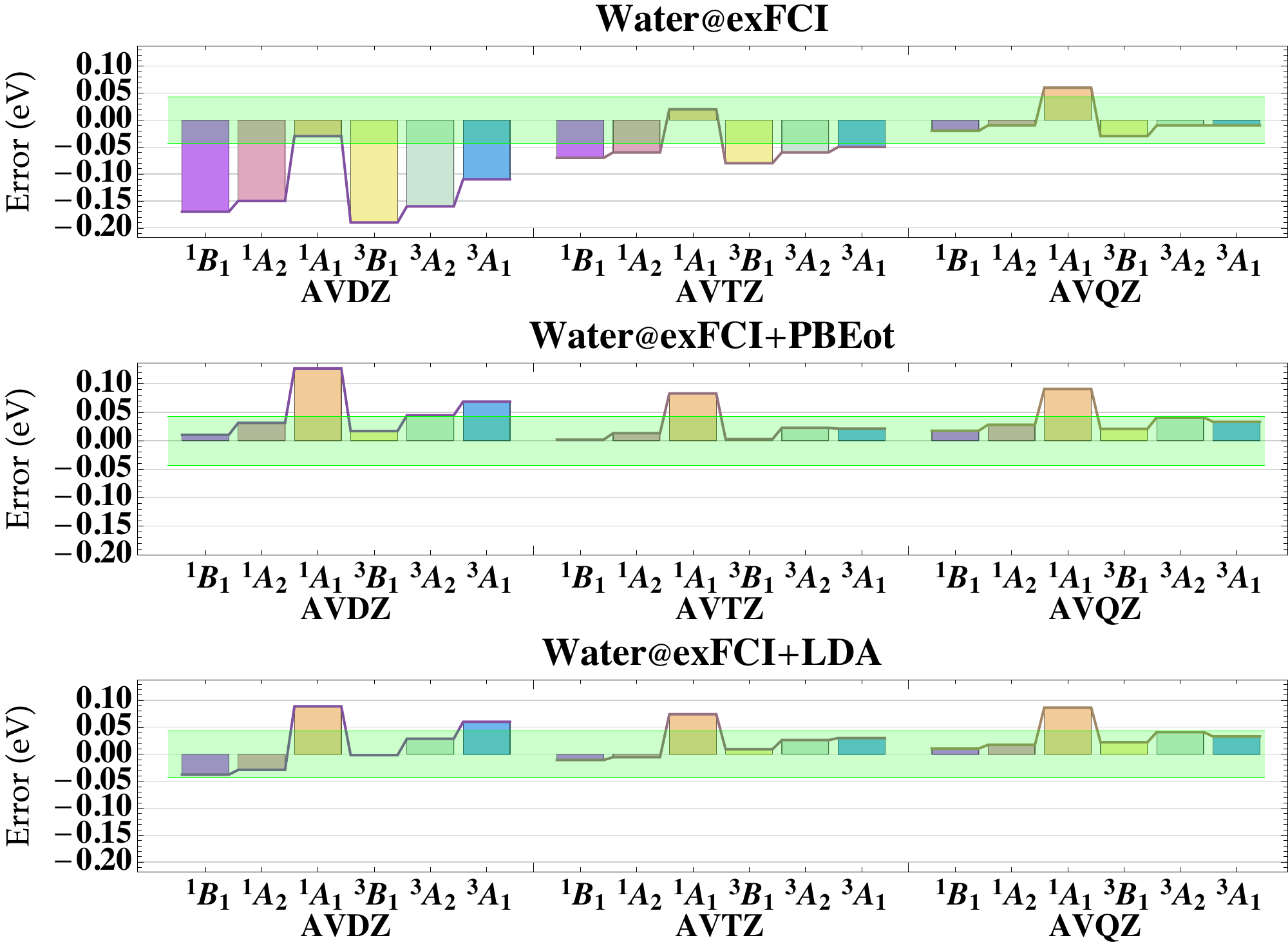}
	\caption{Error in vertical excitation energies (in eV) of water for various basis sets and methods.
	The green region corresponds to chemical accuracy (i.e., error below 1 {\kcal} or 0.043 eV).
	See the {\SI} for raw data.}
	\label{fig:H2O}
\end{figure}

\begin{figure}
	\includegraphics[width=\linewidth]{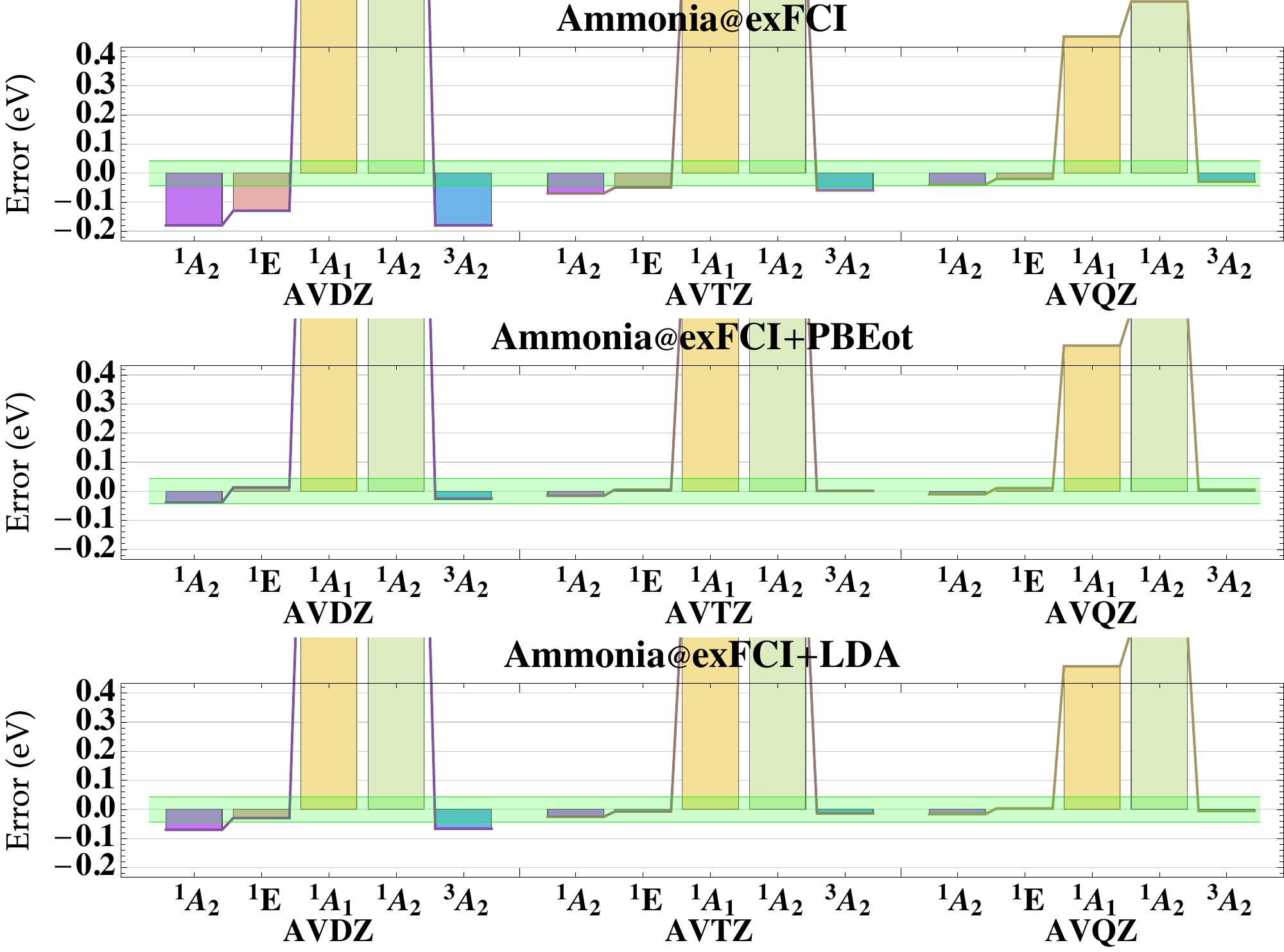}
	\caption{Error in vertical excitation energies (in eV) of ammonia for various basis sets and methods.
	The green region corresponds to chemical accuracy (i.e., error below 1 {\kcal} or 0.043 eV).
	See the {\SI} for raw data.}
	\label{fig:NH3}
\end{figure}

\subsection{Doubly-Excited States of the Carbon Dimer}
\label{sec:C2}
In order to have a miscellaneous test set of excitations, in a third time, we propose to study some doubly-excited states of the carbon dimer \ce{C2}, a prototype system for strongly correlated and multireference systems. \cite{AbrShe-JCP-04, AbrShe-CPL-05, Var-JCP-08, PurZhaKra-JCP-09, AngCimPas-MP-12, BooCleThoAla-JCP-11, Sha-JCP-15, SokCha-JCP-16, HolUmrSha-JCP-17, VarRoc-PTRSMPES-18}
These two valence excitations --- $1\,^{1}\Sigma_g^+ \ra 1\,^{1}\Delta_g$ and $1\,^{1}\Sigma_g^+ \ra 2\,^{1}\Sigma_g^+$ --- are both of $(\pi,\pi) \ra (\si,\si)$ character.
They have been recently studied with state-of-the-art methods, and have been shown to be ``pure'' doubly-excited states as they involve an insignificant amount of single excitations. \cite{LooBogSceCafJac-JCTC-19} 
The vertical excitation energies associated with these transitions are reported in Table \ref{tab:Mol} and represented in Fig.~\ref{fig:C2}.
An interesting point here is that one really needs to consider the $\PBEot$ functional to get chemically accurate excitation energies with the AVDZ atomic basis set.
We believe that the present result is a direct consequence of the multireference character of the \ce{C2} molecule.
In other words, the UEG on-top pair density used in the $\LDA$ and $\PBEUEG$ functionals (see Sec.~\ref{sec:func}) is a particularly bad approximation of the true on-top pair density for the present system.

\begin{figure}
	\includegraphics[width=\linewidth]{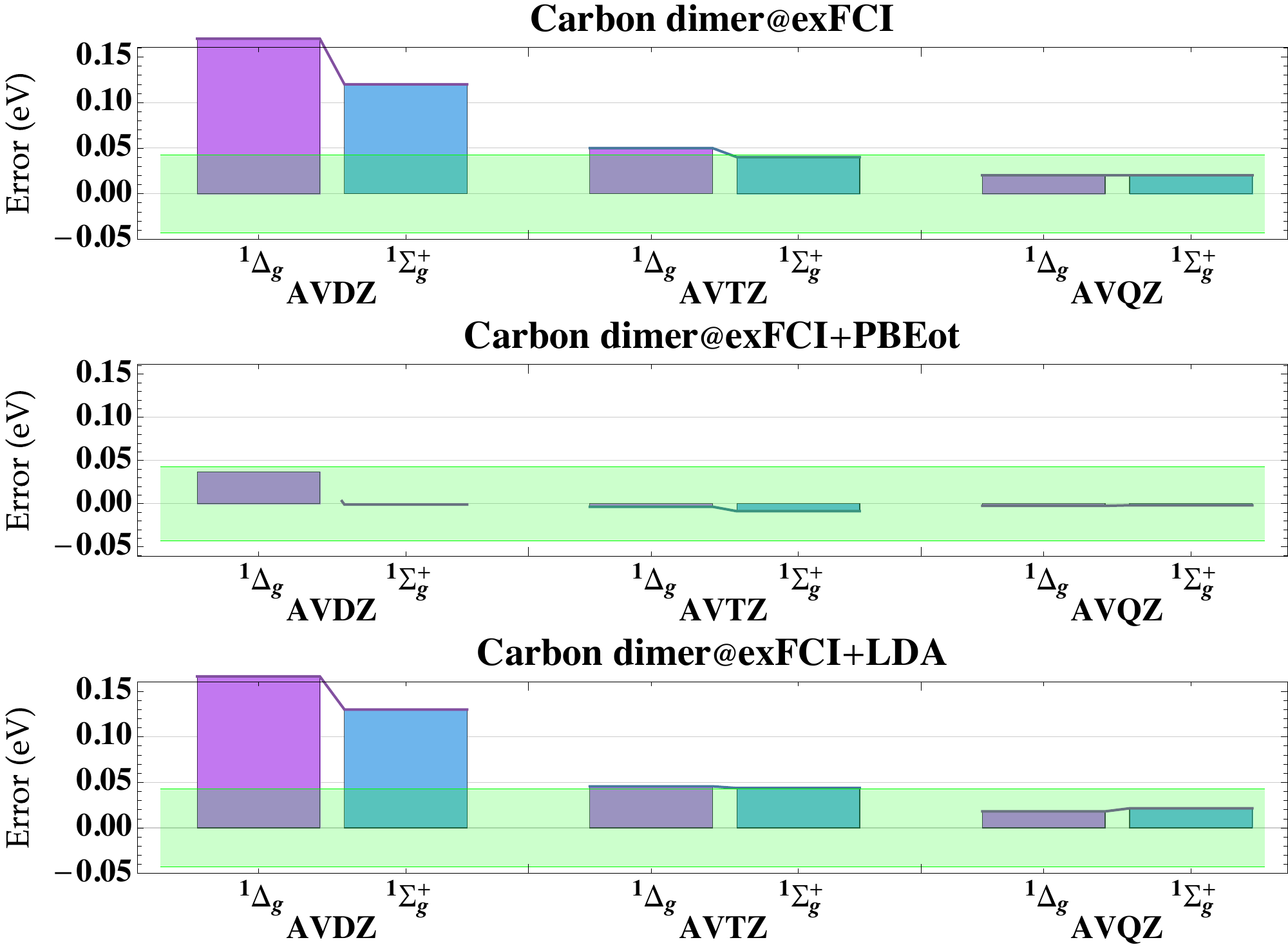}
	\caption{Error in vertical excitation energies (in eV) for two doubly-excited states of the carbon dimer for various basis sets and methods.
	The green region corresponds to chemical accuracy (i.e., error below 1 {\kcal} or 0.043 eV).
	See the {\SI} for raw data.}
	\label{fig:C2}
\end{figure}

It is interesting to study the behavior of the key quantities involved in the basis-set correction for different states as the basis-set incompleteness error is obviously state specific. 
In Fig.~\ref{fig:C2_mu}, we report $\rsmu{}{\Bas}(z)$, $\n{}{\Bas}(z) \be{\text{c,md}}{\sr,\PBEot}(z)$, and $\n{2}{\Bas}(z)$ 
along the nuclear axis ($z$) for the two $^1 \Sigma_g^+$ electronic states of \ce{C2} computed with the AVDZ, AVTZ, and AVQZ basis sets. 
The graphs gathered in Fig.~\ref{fig:C2_mu} illustrate several general features regarding the present basis-set correction: 
\begin{itemize}
	\item the maximal values of $\rsmu{}{\Bas}(z)$ are systematically close to the nuclei, a signature of the atom-centered basis set;
	\item the overall magnitude of $\rsmu{}{\Bas}(z)$ increases with the basis set, which reflects the improvement of the description of the correlation effects when enlarging the basis set;
	\item the absolute value of the energetic correction decreases when the size of the basis set increases;
	\item there is a clear correspondence between the values of the energetic correction and the on-top pair density. 
\end{itemize}
Regarding now the differential effect of the basis-set correction in the special case of the two $^1 \Sigma_g^+$ states studied here, we observe that: 
\begin{itemize}
	\item $\rsmu{}{\Bas}(z)$ has the same overall behavior for the two states, with slightly more fine structure in the case of the ground state. 
	Such feature is consistent with the fact that the two states considered are both of $\Sigma_g^+$ symmetry and of valence character. 
	\item $\n{2}{}(z)$ is overall larger in the excited state, specially in the bonding and outer regions. 
	This is can be explained by the nature of the electronic transition which qualitatively corresponds to a double excitation from $\pi$ to $\sigma$ orbitals, therefore increasing the overall electronic population on the bond axis. 
	\item The energetic correction clearly stabilizes preferentially the excited state rather than the ground state, illustrating that short-range correlation effects are more pronounced in the former than in the latter. 
	This is linked to the larger values of the excited-state on-top pair density. 
\end{itemize}

\begin{figure*}
	\includegraphics[height=0.35\linewidth]{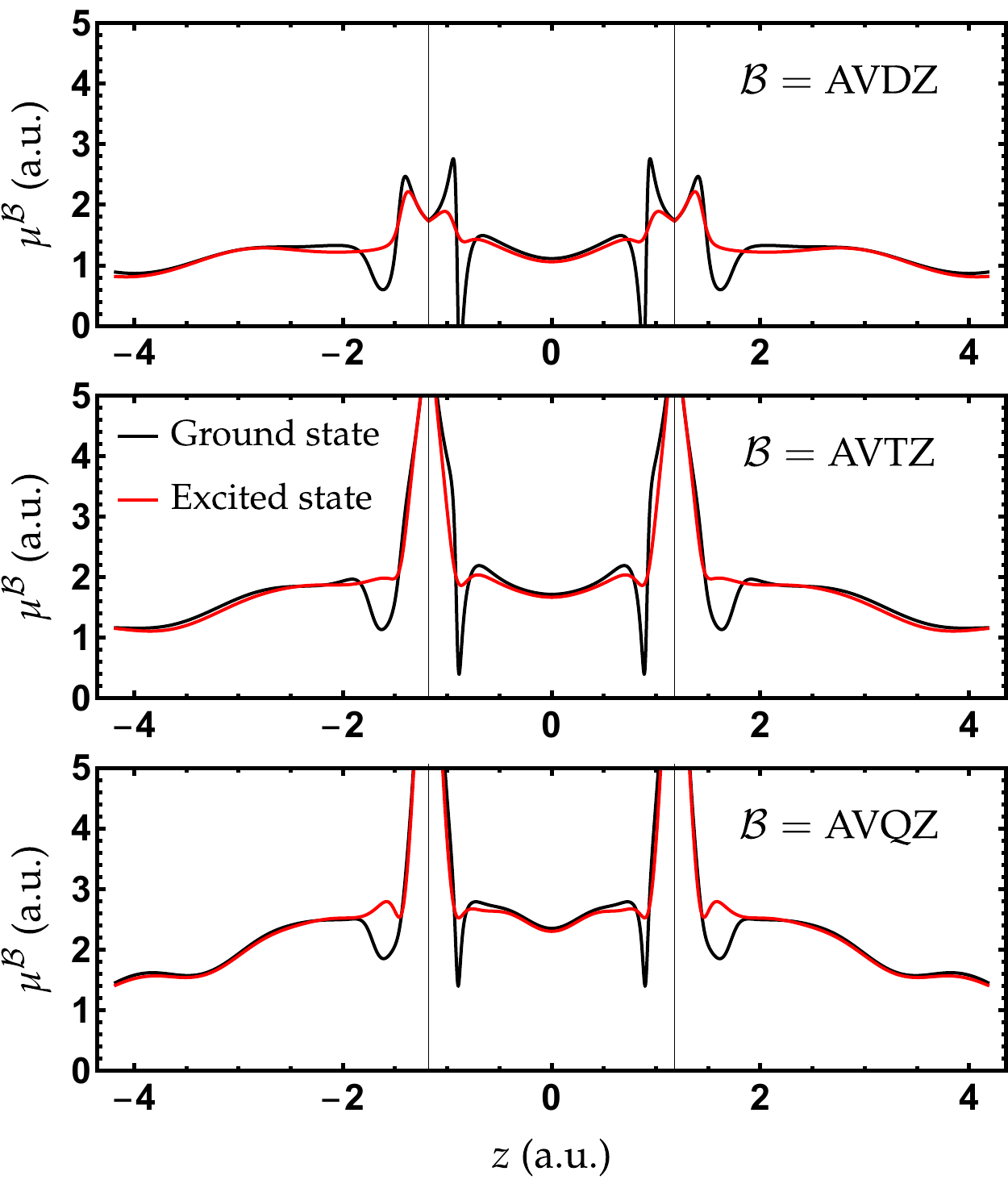}
	\includegraphics[height=0.35\linewidth]{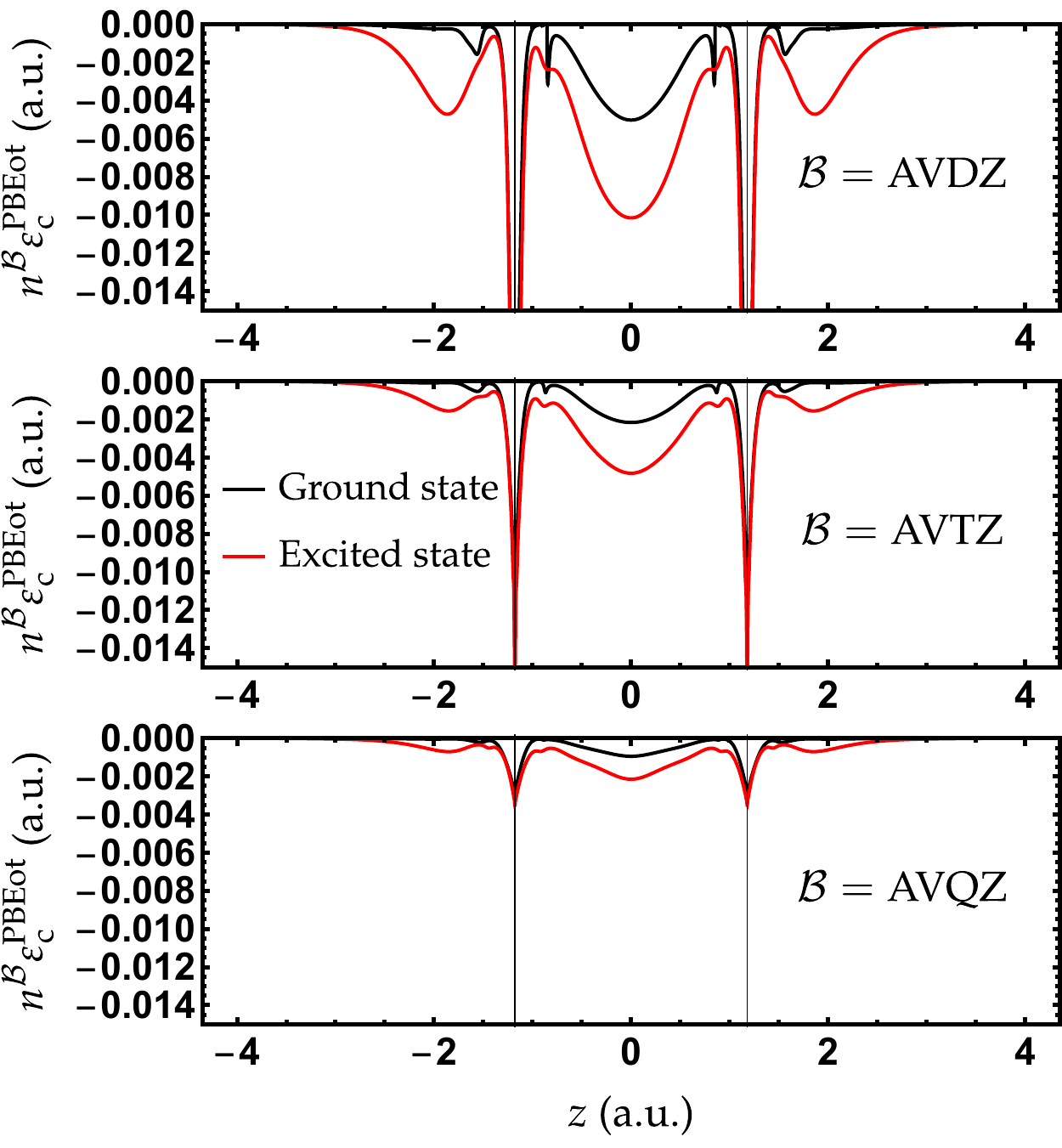}
	\includegraphics[height=0.35\linewidth]{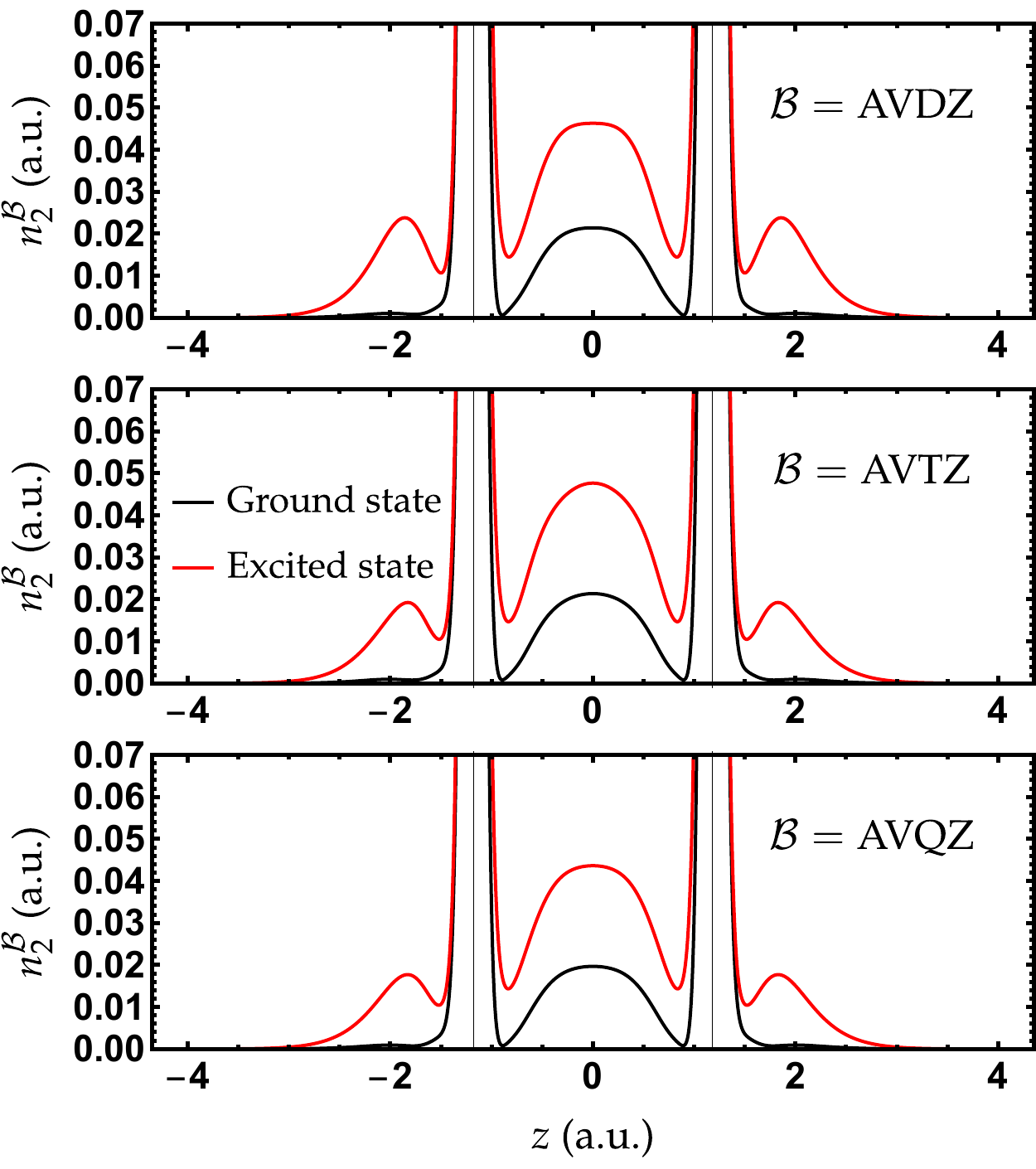}
	\caption{$\rsmu{}{\Bas}$ (left), $\n{}{\Bas} \be{\text{c,md}}{\sr,\PBEot}$ (center) and $\n{2}{\Bas}$ (right) along the molecular axis ($z$) for the ground state (black curve) and second doubly-excited state (red curve) of \ce{C2} for various basis sets $\Bas$.
	The two electronic states are both of $\Sigma_g^+$ symmetry. 
	The carbon nuclei are located at $z= \pm 1.180$ bohr and are represented by the thin black lines.}

	\label{fig:C2_mu}
\end{figure*}

\subsection{Ethylene}
\label{sec:C2H4}

As a final example, we consider the ethylene molecule, yet another system which has been particularly scrutinized theoretically using high-level ab initio methods. \cite{SerMarNebLinRoo-JCP-93, WatGwaBar-JCP-96, WibOliTru-JPCA-02, BarPaiLis-JCP-04, Ang-JCC-08, SchSilSauThi-JCP-08, SilSchSauThi-JCP-10, SilSauSchThi-MP-10, Ang-IJQC-10, DadSmaBooAlaFil-JCTC-12, FelPetDav-JCP-14, ChiHolAdaOttUmrShaZim-JPCA-18}
We refer the interested reader to the work of Feller \textit{et al.}\cite{FelPetDav-JCP-14} for an exhaustive investigation dedicated to the excited states of ethylene using state-of-the-art CI calculations.
In the present context, ethylene is a particularly interesting system as it contains a mixture of valence and Rydberg excited states. 
Our basis-set corrected vertical excitation energies are gathered in Table \ref{tab:Mol} and depicted in Fig.~\ref{fig:C2H4}.
The exFCI+$\PBEot$/AVDZ excitation energies are at near chemical accuracy and the errors drop further when one goes to the triple-$\zeta$ basis.
Consistently with the previous examples, the $\LDA$ and $\PBEUEG$ functionals are slightly less accurate, although they still correct the excitation energies in the right direction.

\begin{figure}
	\includegraphics[width=\linewidth]{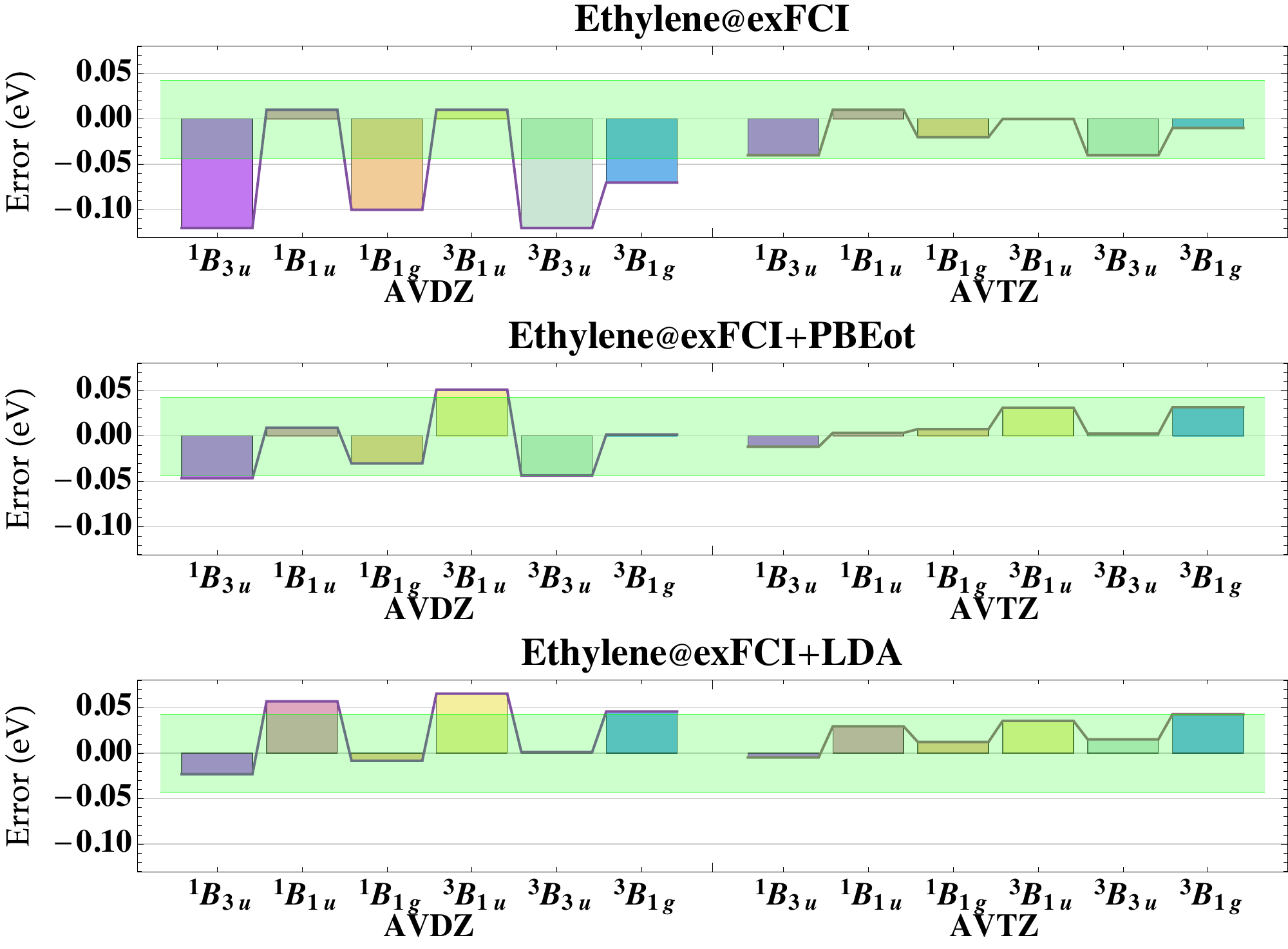}
	\caption{Error in vertical excitation energies (in eV) of ethylene for various basis sets and methods.
	The green region corresponds to chemical accuracy (i.e., error below 1 {\kcal} or 0.043 eV).
	See the {\SI} for raw data.}
	\label{fig:C2H4}
\end{figure}

\section{Conclusion}
\label{sec:ccl}
We have shown that, by employing the recently proposed density-based basis-set correction developed by some of the authors, \cite{GinPraFerAssSavTou-JCP-18} one can obtain, using sCI methods, chemically accurate excitation energies with typically augmented double-$\zeta$ basis sets.
This nicely complements our recent investigation on ground-state properties, \cite{LooPraSceTouGin-JPCL-19} which has evidenced that one recovers quintuple-$\zeta$ quality atomization and correlation energies with triple-$\zeta$ basis sets.
The present study clearly shows that, for very diffuse excited states, the present correction relying on short-range correlation functionals from RS-DFT might not be enough to catch the radial incompleteness of the one-electron basis set.
Also, in the case of multireference systems, we have evidenced that the $\PBEot$ functional, which uses an accurate on-top pair density, is more appropriate than the $\LDA$ and $\PBEUEG$ functionals relying on the UEG on-top pair density.
We are currently investigating the performance of the present basis-set correction for strongly correlated systems and we hope to report on this in the near future.

\section*{Supporting Information Available}
See {\SI} for geometries and additional information (including total energies and energetic correction of the various functionals).

\begin{acknowledgements}
PFL would like to thank Denis Jacquemin for numerous discussions on excited states.
This work was performed using HPC resources from GENCI-TGCC (Grant No.~2018-A0040801738), CALMIP (Toulouse) under allocation 2019-18005 and the Jarvis-Alpha cluster from the \textit{Institut Parisien de Chimie Physique et Th\'eorique}.
\end{acknowledgements}

\bibliography{Ex-srDFT,Ex-srDFT-control}

\end{document}